\documentclass{PoS}

\newcommand{\ket}[1]{| #1 \rangle}
\newcommand{\nf}{{N_\mathrm{f}}}
\newcommand{\mK}{m_\mathrm{K}}
\newcommand{\mpi}{m_\pi}
\newcommand{\mev}{\,\mathrm{MeV}}
\newcommand{\fm}{\,\mathrm{fm}}
\newcommand{\gev}{\,\mathrm{GeV}}

\newcommand{\eq}[1]{Eq.~(\ref{#1})}
\newcommand{\fig}[1]{Fig.~\ref{#1}}

\title{Testing the hadro-quarkonium model on the lattice}

\ShortTitle{Hadro-quarkonium}

\author{\speaker{Francesco Knechtli}, Maurizio Alberti\\
        Department of Physics, Bergische Universit\"at Wuppertal\\
        Gau\ss{}stra\ss{}e 20, 42119 Wuppertal, Germany\\
        E-mail: \email{knechtli@physik.uni-wuppertal.de},
                \email{alberti@uni-wuppertal.de}}

\author{Gunnar S.~Bali, Sara Collins, Wolfgang S\"oldner\\
        Institut f\"ur Theoretische Physik, Universit\"at Regensburg\\
        Universit\"atsstra\ss{}e 31, 93053 Regensburg, Germany\\
        E-mail: \email{gunnar.bali@ur.de},
                \email{sara.collins@ur.de},
                \email{wolfgang.soeldner@physik.uni-regensburg.de}}

\author{Graham Moir\\
        Department of Applied Mathematics and Theoretical Physics, Centre for Mathematical Sciences\\
        University of Cambridge, Wilberforce Road, Cambridge, CB3 0WA, UK\\
        E-mail: \email{graham.moir@damtp.cam.ac.uk}}

\abstract{Recently the LHCb experiment found evidence for the existence of two
exotic resonances consisting of $c\bar{c}uud$ quarks. Among the possible
interpretations is the hadro-charmonium model, in which charmonium
is bound ``within" a light hadron. We test this idea on CLS $\nf$=2+1 lattices
using the static formulation for the heavy quarks. We find that the
static potential is modified by the presence of a hadron such that it becomes
more attractive. The effect is of the order of a few MeV.
\begin{flushright} WUB/16-09\\
\end{flushright}
}

\FullConference{34th Annual International Symposium on Lattice Field Theory\\
		24--30 July 2016\\
		University of Southampton, UK}

\begin{document}

\section{Motivation}

The LHCb collaboration analysed the decay $\Lambda_b\rightarrow J/\psi~p~K$
\cite{Aaij:2015tga,Aaij:2016phn}.
A satisfactory description of the data is obtained by adding to the 
$\Lambda^*\rightarrow p~K$ resonances two additional 
resonances of exotic quark content $uudc\bar{c}$ labeled by 
$P_{c}^{+}(4380)$ $(J^{P}=\frac{3}{2}^{-})$ and 
$P_{c}^{+}(4450)$ $(J^{P}=\frac{5}{2}^{+})$
and decaying strongly into $J/\psi~p$. Flipping the two parities
can also explain the data~\cite{Aaij:2015tga,Aaij:2016phn}.
Attractive forces between charmonium and $pp$ systems have been previously
conjectured, e.g.,
to explain the rapid change in the behavior of polarized $pp$ 
scattering around $\sqrt{s}=5\gev\approx m_{p}+m_{p}+m_{J/\psi}$ \cite{Brodsky:1989jd}.

Five (4 $q$, 1 $\bar{q}$) quark systems are very difficult to study
directly on the lattice. For example see \cite{Beane:2014sda}
for a study of a charmonium-nucleon system.
Here we test a particular model instead, hadro-quarkonium. In this model
quarkonia are bound ``within'' ordinary hadrons \cite{Dubynskiy:2008mq}.
Examples of charmonium-baryon systems which are close in energy to the
LHCb pentaquark candidates are
$m(\Delta)+m(J/\psi)\approx4329\mev$ for $J^P=\frac{3}{2}^{-}$ and
$m(N)+m(\chi_{c2})\approx4496\mev$ for $J^P=\frac{5}{2}^{+}$.

\section{Hadro-quarkonia in the static limit}

The hadro-quarkonium model can be tested in the static quark limit.
To leading order in potential non-relativistic QCD, quarkonia
can be approximated by the non-relativistic Schr\"odinger
equation with a static quark-antiquark potential $V_0(r)$. The question 
we want to
answer in our study \cite{Alberti:2016dru} is whether the static potential
becomes more or less attractive, when light hadrons are ``added''.
For this we create a zero-momentum projected
hadronic state $\ket{H}$ at the time $0$. 
We let it propagate for an interval $\delta t$ and we then create a 
quark-antiquark ``string''. They propagate together for a time interval $t$.
We destroy the string at time $t+\delta t$ and finally the light hadron at time
$t+2\delta t$.
We compute the correlator
\begin{equation}\label{eq:correlator}
C_H(r, \delta t, t) =
\frac{\langle W(r,t)C_{H,\mathrm{2pt}}(t+2\delta t)\rangle}{\langle W(r,t)\rangle
\langle C_{H,\mathrm{2pt}}(t+2\delta t)\rangle}\,,
\end{equation}
where we average over the spatial positions of the Wilson loop $W(r,t)$
and over the hadronic sink positions in the hadronic two-point function
$C_{H,\mathrm{2pt}}$.
The difference between the static potential in the presence of a
hadron $V_H$ and the potential in the vacuum $V_0$
can be obtained from
\begin{equation}\label{eq:difference}
\Delta V_H(r,\delta t) \equiv
V_H(r,\delta t)-V_0(r)
=-\lim_{t\rightarrow\infty}\frac{{\rm d}}{{\rm d}t}\ln[C_H(r, \delta t, t)]
\end{equation}
and extrapolating $\delta t\rightarrow\infty$.

\section{Lattice results}

We analyse the $\nf=2+1$ CLS ensemble ``C101'' which has 
$96\times 48^3$ sites,
$\mpi=220\mev$, $\mK=470\mev$, 
$L\mpi\approx 4.6$, $L\approx 4.1\fm$,
$t_0/a^2=2.9085(51)$ \cite{Bruno:2014jqa}.
It has been simulated using the publicly available openQCD package \cite{algo:openQCD}.
The lattice spacing $a=0.0854(15)\fm$ is determined 
from the scale $\sqrt{8t_0}/a$ \cite{flow:ML}
extrapolated to physical point \cite{Bali:2016umi}
and using $\sqrt{8t_0}=0.4144(59)(37)\fm$ \cite{Borsanyi:2012zs}.
We perform a large statistics calculation consisting
of 1552 configurations separated by
4 MDUs, times 12 hadron sources
(providing 10 forward and backward propagating two-point functions and for the two sources closest
to the open temporal boundaries a forward and a backward two-point function for a total of 22 correlation functions).
Wilson loops are measured at all positions and in each direction separately.
Hadronic two-point functions and Wilson loops are smeared to optimize
their overlap with the respective ground states.
We measure $\Delta V_H$ for $\pi$, $K$, $\rho$, $K^{\star}$ and $\phi$ mesons;
for $N$, $\Sigma$, $\Lambda$ and $\Xi$ octet baryons with $J^P=\frac{1}{2}^{\pm}$;
 and for $\Delta$, $\Sigma^{\star}$, $\Xi^{\star}$ and $\Omega$ decuplet baryons
with $J^P=\frac{3}{2}^{\pm}$.
\begin{figure}[t]\centering
  \resizebox{10cm}{!}{\includegraphics{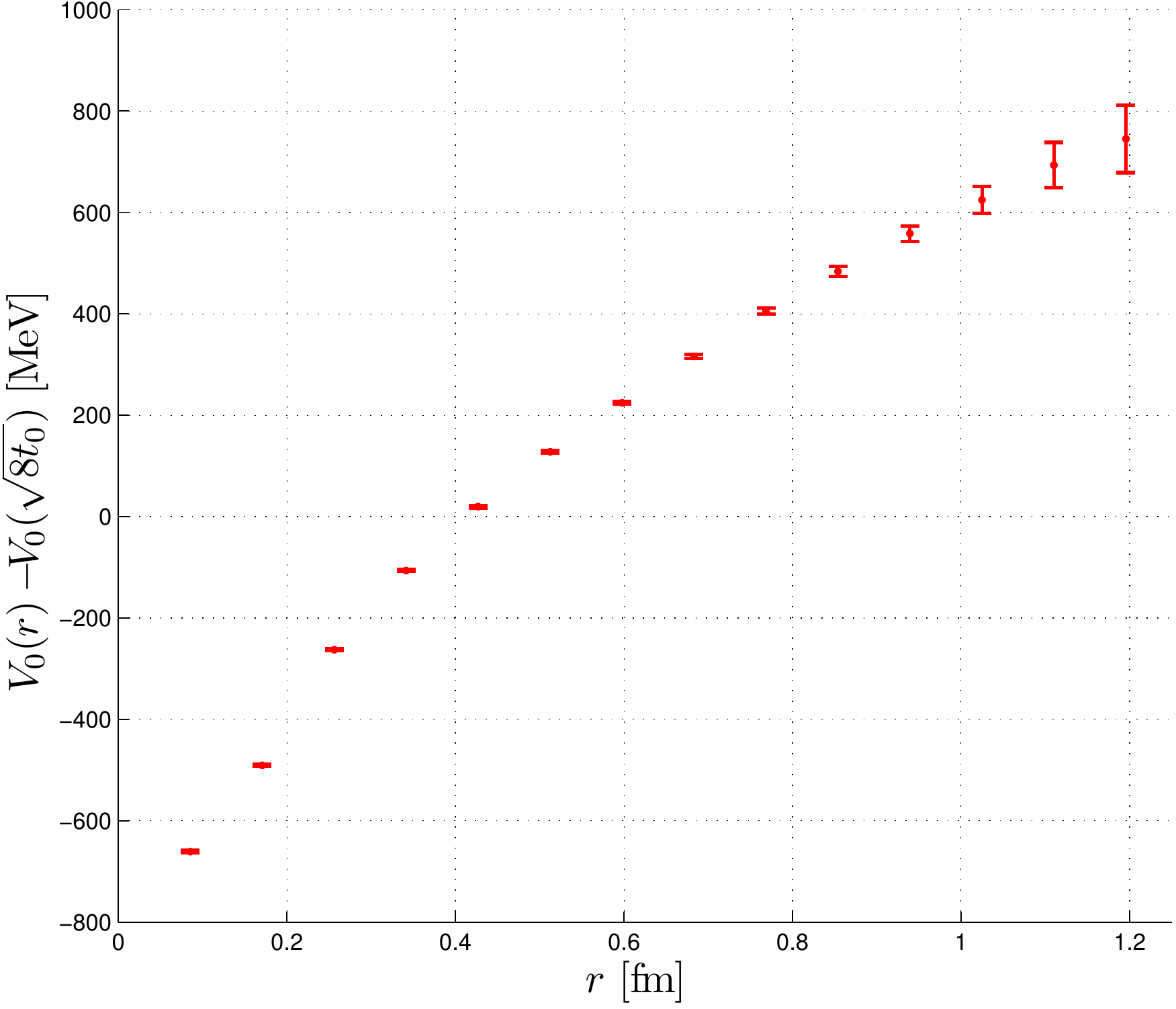}}
  \caption{The static potential in the vacuum measured on the CLS ensemble
``C101''.}
  \label{f:potential}
\end{figure}

We begin the presentation of the results by showing in \fig{f:potential}
the static potential $V_0(r)$ in the vacuum. It has been determined using 
the methods of \cite{Donnellan:2010mx}. We plot $V_0(r)$ for distances 
$r\le1.2\fm$ below the string breaking region.

\begin{figure}[t]\centering
  \resizebox{10cm}{!}{\includegraphics{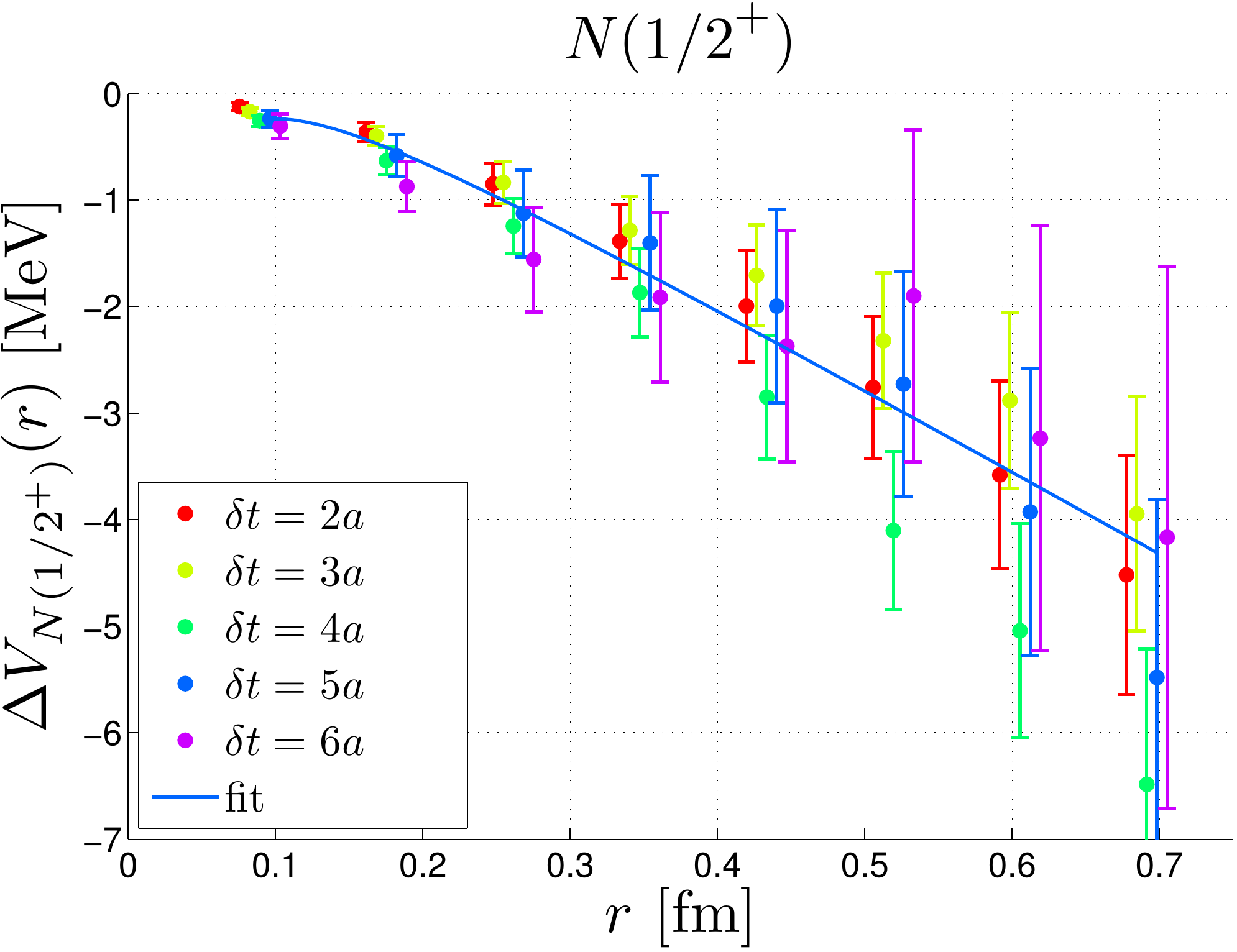}}
  \caption{Modification of the static potential ``within'' a nucleon $N(\frac{1}{2}^+)$.}
  \label{f:nucleon}
\end{figure}

\begin{figure}[t]\centering
  \resizebox{10cm}{!}{\includegraphics{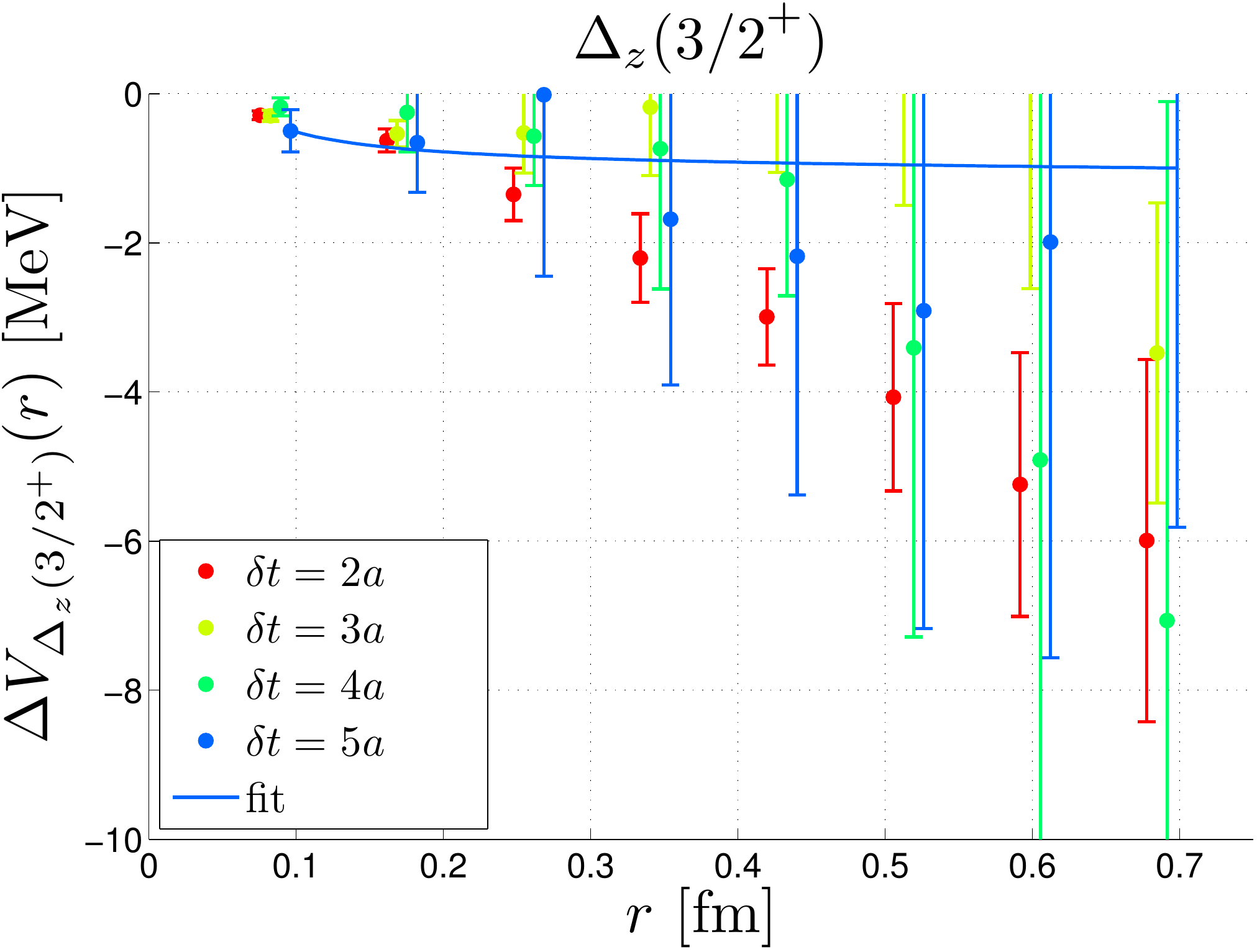}}
  \caption{Modification of the static potential ``within'' a $\Delta(\frac{3}{2}^{+})$.}
  \label{f:delta}
\end{figure}

\begin{figure}[t]\centering
  \resizebox{10cm}{!}{\includegraphics{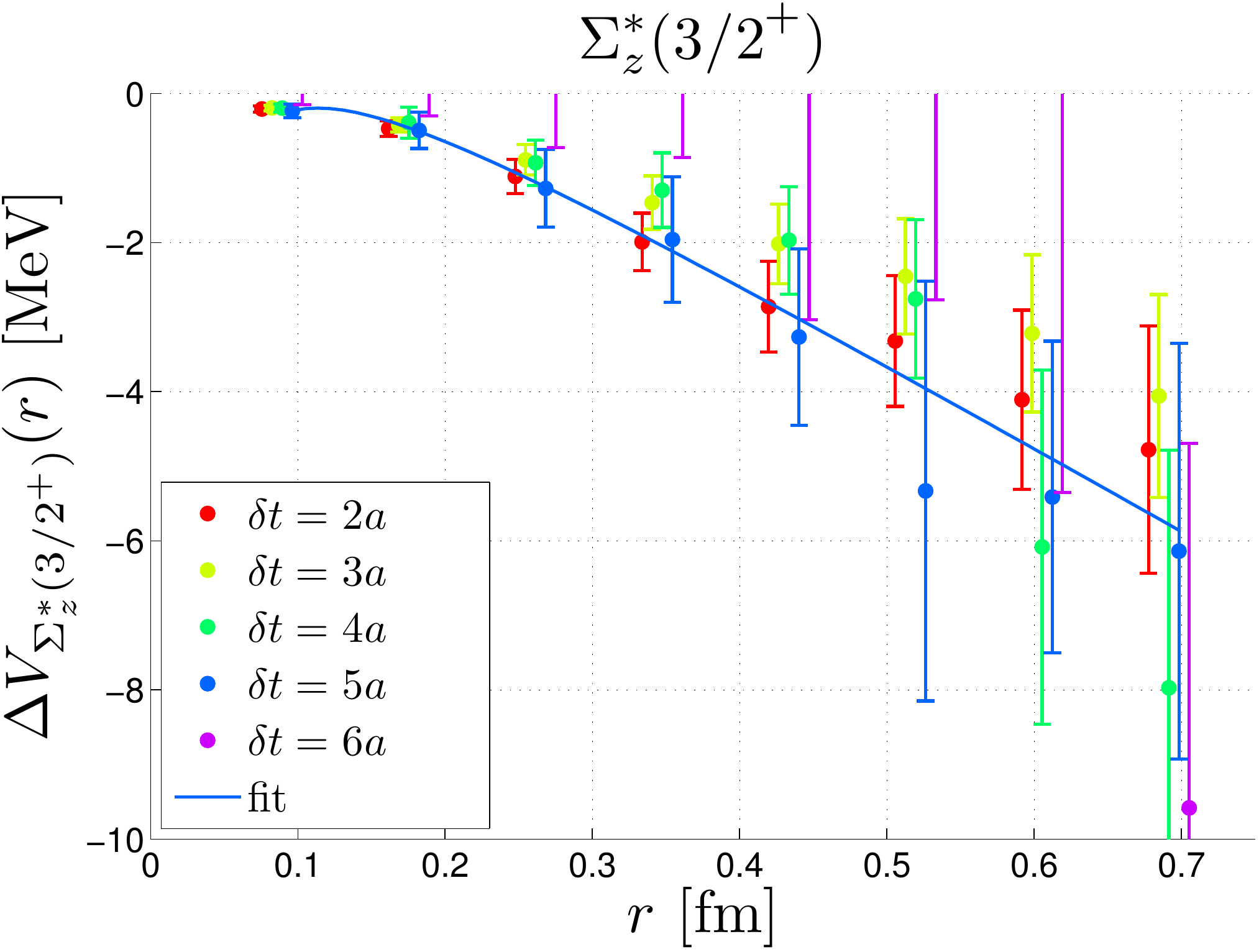}}
  \caption{Modification of the static potential ``within'' a $\Sigma^{*}(\frac{3}{2}^{+})$.}
  \label{f:sigmastar_p}
\end{figure}

\begin{figure}[t]\centering
  \resizebox{10cm}{!}{\includegraphics{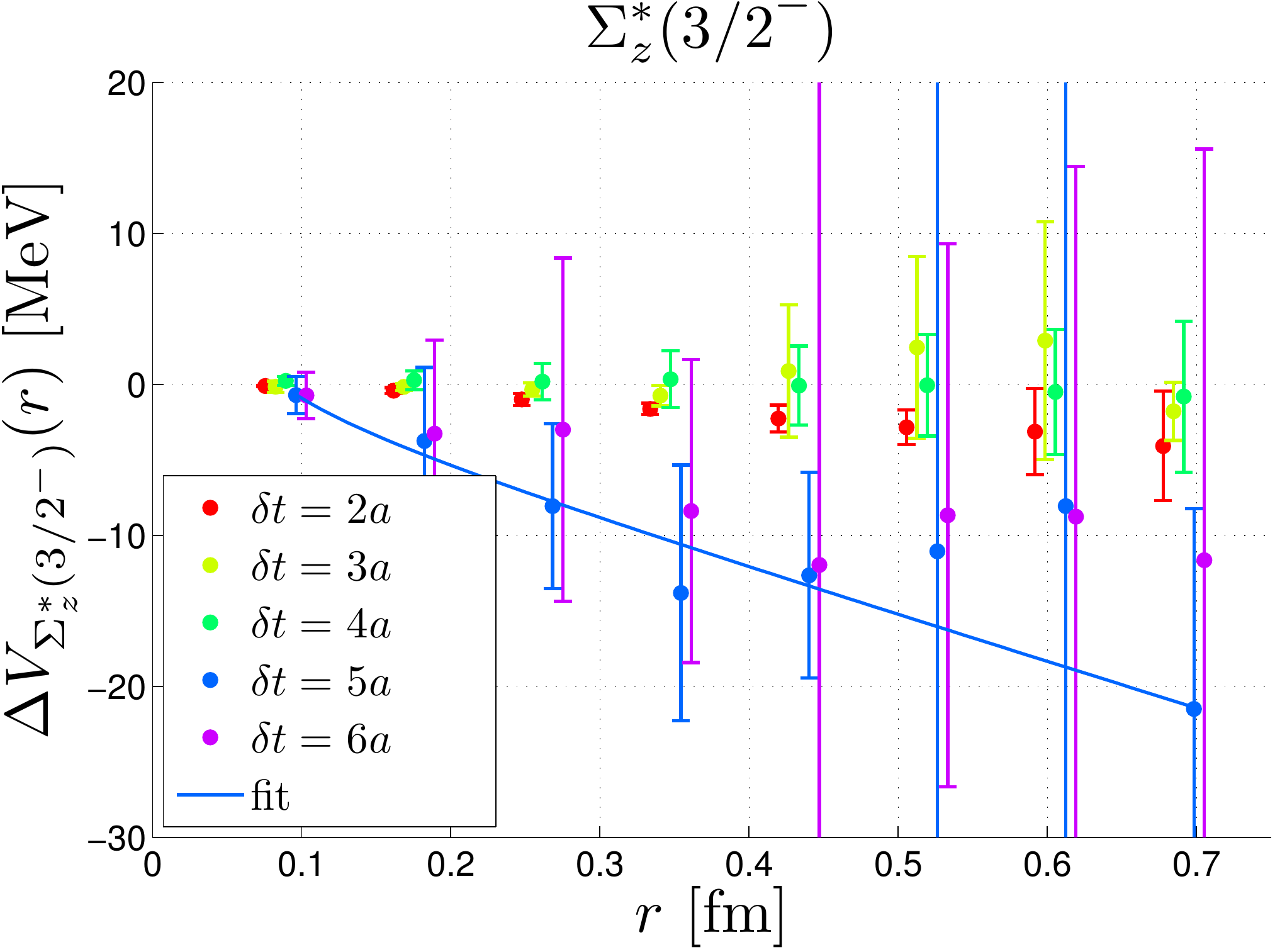}}
  \caption{Modification of the static potential ``within'' a $\Sigma^{*}(\frac{3}{2}^{-})$.}
  \label{f:sigmastar_n}
\end{figure}

In order to extract the energy difference $\Delta V_H$ in \eq{eq:difference}
for a given hadron labeled by $H(J^P)$,
for each combination of $r$ and $\delta t$, we perform
linear fits in $t$ to $\ln[C_H(r,\delta t,t)]$.
The range of $t$ for the fits is chosen in the region where the
effective energy $a^{-1}\ln[C_H(r, \delta t, t)/C_H(r, \delta t, t+a)]$
exhibits a clear plateau. In \fig{f:nucleon}, \fig{f:delta}, \fig{f:sigmastar_p}
and \fig{f:sigmastar_n} we show the results for the positive parity nucleon,
$\Delta$ and $\Sigma^*$ as well as for the negative parity $\Sigma^*$,
respectively.
Notice that different colors in the plots correspond to different values of 
$\delta t$ which are slightly displaced horizontally for clarity.
We display the statistical errors only. In \cite{Alberti:2016dru} we also
give estimates of the systematic error by changing the range of the fits.

In \fig{f:nucleon} we show $\Delta V_H(r,\delta t)$ for
the nucleon $N(\frac{1}{2}^+)$.
We observe $\Delta V_H(r,\delta t)<0$.
The results agree for $\delta t\gtrsim 3a$ and we take the values
for $\delta t=5a$ as good approximation to the limit $\delta t\to\infty$.
The data are well described by a fit to the Cornell parametrization
\begin{equation}\label{eq:parametrization}
\Delta V_H(r,\delta t=5a) = \Delta \mu_H - \frac{\Delta c_H}{r} + \Delta \sigma_H r
\end{equation}
with the parameters $\Delta \mu_H$, $\Delta c_H$ and $\Delta \sigma_H$,
also shown in \fig{f:nucleon}.
We find that the size of the effect is $\Delta V_H(r)\approx-1\mev$ to 
$-2\mev$ at a distance $r\simeq0.3\fm$ and grows to 
$\Delta V_H(r)\approx-4\mev$ to $-7\mev$ at
our largest shown distance $r\simeq0.7\fm$. Notice that a bound state of the
nucleon $N(\frac{1}{2}^+)$ with a $\chi_{c2}(2^{+})$ could explain the 
$J^P=\frac{5}{2}^{+}$ pentaquark resonance.

In \fig{f:delta} we show $\Delta V_H(r,\delta t)$ for
the $\Delta(\frac{3}{2}^{+})$. In this case, in \eq{eq:correlator} we
correlate the $\Delta$ polarized in $z$ direction with Wilson loops taken
in $z$ direction only, to guarantee that we project onto spin 
$\Lambda=|J_z|=3/2$ along the distance in $z$-direction between the static 
sources. We find similar results as for the nucleon, albeit with rather
large errors. Notice that a bound state of a $\Delta(\frac{3}{2}^{+})$ with a 
$J/\psi(1^{-})$ could explain the $J^P=\frac{3}{2}^{-}$ pentaquark resonance.
As another example with the same spin and parity assignment but with a 
strange quark content, in \fig{f:sigmastar_p} we show $\Delta V_H(r,\delta t)$
for the decuplet $\Sigma^{*}(\frac{3}{2}^{+})$. The results are very
similar to those for the nucleon.

In \fig{f:sigmastar_n} we show an example of $\Delta V_H(r,\delta t)$ for
a negative parity state, the decuplet $\Sigma^{*}(\frac{3}{2}^{-})$. 
The statistical errors are much larger than for the positive parity case 
shown in \fig{f:sigmastar_p}. Within the errors the values of $\Delta V_H$
are consistent with the positive parity case but even larger negative values 
cannot be excluded for the negative parity case. Notice that a bound state of a 
$\Sigma^{*}(\frac{3}{2}^{-})$ with a $J/\psi(1^{-})$ could give a
$J^P=\frac{5}{2}^{+}$ pentaquark resonance. However in this case it contains a
strange quark and also the resulting
mass is too large and does not match the mass of the $P_{c}^{+}(4450)$ pentaquark.

\section{Volume check}

\begin{figure}[t]\centering
  \resizebox{8cm}{!}{\includegraphics{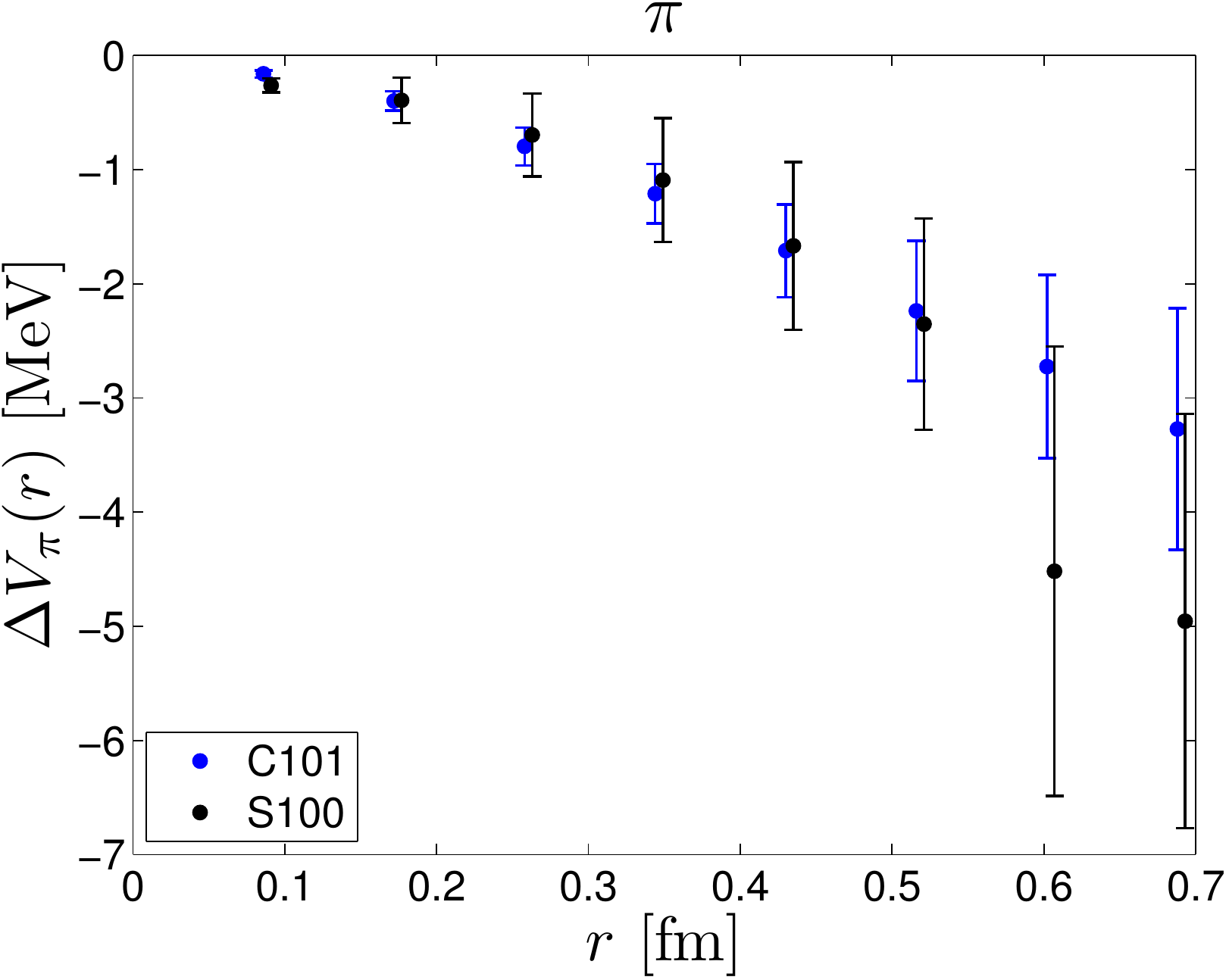}}
  \caption{Comparison of the modification of the static potential
``within'' a pion for two different volumes.}
  \label{f:volume}
\end{figure}
In order to check for finite volume effects we analysed a second CLS ensemble
``S100'' with a smaller volume of $128\times32^3$ sites but with the same 
lattice spacing and quark masses as the ensemble ``C101''.
At the time of the lattice conference the statistics were insufficient to draw
a conclusion. By the time these proceedings were written,
data for $478$ configurations of ``S100'' were available. 
In \fig{f:volume} we compare the results for $\Delta V_{\pi}(r,5a)$ 
on the ``C101'' and ``S100'' ensembles and the
conclusion is that there are no significant finite volume effects.

\section{Conclusions}

We have numerically established the modification $\Delta V_H$ of the static 
quark-antiquark potential in the presence of a hadron, see \eq{eq:difference}.
We find
$\Delta V_H(r)<0$. At a distance of $0.5\fm$ the size of the effect varies
between $2\mev$ and $3\mev$ for all the hadrons we investigated. 
The main effect can be parametrized as a reduction of the linear slope of 
the static potential. We emphasize that we do not see finite volume effects,
comparing $Lm_{\pi}\approx 4.6$ (``C101'') with $Lm_{\pi}\approx 3.1$ (``S100'').

In order to answer the question, whether this modification leads to a
larger binding energy of charmonium states, we have compared the energy
levels that result from solving the Schr\"odinger equation with the
vacuum static potential $V_0$ and with the modified potential 
$V_0+\Delta V_H$. Details of this calculation can be found in 
\cite{Alberti:2016dru}. The result is a stronger binding of charmonium 
$1S$ state by $-1\mev$ to $-2.5\mev$,
of $1P$ state by $-1\mev$ to $-5\mev$
and of $2S$ state by $-1\mev$ to $-6.5\mev$.
These binding energies are similarly small in size as in the deuterium
system and may be
somewhat inconsistent with the original hadro-charmonium picture.

{\bf Acknowledgments.}
This work was supported by the Deutsche Forschungsgemeinschaft Grant
No.\ SFB/TRR 55. GM acknowledges support from the Herchel Smith Fund at
the University of Cambridge and the Deutsche Forschungsgemeinschaft (DFG) 
under contract KN 947/1-2.
We acknowledge PRACE for awarding CLS access to resources 
on Fermi at CINECA Bologna and 
on SuperMUC at Leibniz Supercomputing Centre Munich.
We also acknowledge computer time granted 
on the ``Clover'' Cluster of the Mainz Helmholtz Institute for the
ensemble generation and
on the SFB/TRR~55 QPACE~2 Xeon-Phi installation at Regensburg
and on the Stromboli cluster in Wuppertal for the measurements.
The calculation of hadronic two point-functions is based on the
{\tt CHROMA}~\cite{Edwards:2004sx} software package.
Wilson loops are computed using B.~Leder's
program available at {\tt https://github.com/bjoern-leder/wloop/}.
For the error analysis we applied the method of \cite{Wolff:2003sm}
including the reweighting factors, see \cite{algo:openQCD}.

\end{document}